# Graphene-Complex-Oxide Nanoscale Device Concepts

Giriraj Jnawali,[1,2] Hyungwoo Lee,[3] Jung-Woo Lee,[3] Mengchen Huang,[1,2] Jen-Feng Hsu,[1,2] Bi Feng,[1,2] Rongpu Zhou,[1,2] Guanglei Cheng,[1,2] Brian D'Urso,[1,2] Patrick Irvin,[1,2] Chang-Beom Eom,[3] and Jeremy Levy[1,2*]

[1]*Department of Physics and Astronomy, University of Pittsburgh, Pittsburgh 15260, USA*
[2]*Pittsburgh Quantum Institute, Pittsburgh, PA  15260, USA*
[3]*Department of Materials Science and Engineering, University of Wisconsin-Madison, Madison, WI 53706, USA*

## Abstract

The integration of graphene with complex-oxide heterostructures such as $LaAlO_3/SrTiO_3$ offers the opportunity to combine the multifunctional properties of an oxide interface with the electronic properties of graphene. The ability to control interface conduction through graphene and understanding how it affects the intrinsic properties of an oxide interface are critical to the technological development of novel multifunctional devices. Here we demonstrate several device archetypes in which electron transport at an oxide interface is modulated using a patterned graphene top gate. Nanoscale devices are fabricated at the oxide interface by conductive atomic force microscope (c-AFM) lithography, and transport measurements are performed as a function of the graphene gate voltage. Experiments are performed with devices written adjacent to or directly underneath the graphene gate. Unique capabilities of this approach include the ability to create highly flexible device configurations, the ability to modulate carrier density at the oxide interface, and the ability to control electron transport up to the single-electron-tunneling regime, while maintaining intrinsic transport properties of the oxide interface. Our results facilitate the design of a variety of nanoscale devices that combine unique transport properties of these two intimately coupled two-dimensional electron systems.

**KEYWORDS:** Graphene, complex-oxide interfaces, nanoscale devices, top gating, AFM lithography, electron transport

[*] Correspondence and requests for materials should be addressed to J.L. (jlevy@pitt.edu).

The year 2004 marked milestones for research in graphene[1] and complex-oxide heterostructures.[2] The elegant simplicity of graphene,[3,4] combined with the relative ease of creating graphene-based devices via mechanical exfoliation, contrasts with complex oxides, which need to be synthesized using sophisticated growth techniques and which exhibit behavior that is remarkable but far less well understood. Interest in the LaAlO$_3$/SrTiO$_3$ (LAO/STO) system centered around the emergent conductivity of the interface (both LAO and STO are band insulators), but quickly expanded to include many tunable properties including a metal-insulator transition[5] superconductivity,[6,7] magnetism,[8-11] and spin-orbit interactions.[12,13] The development of reconfigurable nanoscale control of these properties using conductive-atomic force microscope (c-AFM) lithography[14,15] has raised hope that rich functionality at the nanoscale may be achievable from oxide nanoelectronic platforms.

One of the most striking properties of LAO/STO heterostructures is the existence of a sharp metal-insulator transition (MIT) at a critical thickness of LAO, $d_c \sim 4$ unit cell (uc). At or above this thickness, the interface is conducting.[5] Just below this thickness, $d \sim 3$ uc, the interface is nominally insulating but can be locally and reversibly switched between conductive and insulating phases using a voltage-biased conductive atomic force microscope (c-AFM) tip.[14,15] By scanning a c-AFM tip in a controlled fashion, one can create rewritable nanoscale interfacial devices. Recent studies have demonstrated that this technique can also be used to create devices on graphene/LAO/STO (G/LAO/STO) heterostructures, providing a crucial breakthrough for fabricating vertical field-effect devices at the interface as well as on graphene.[16] Due to very thin (< 2 nm) separation between graphene and the oxide interface, strong proximity-induced electronic coupling and correlations are expected. Recently, suppressed electron-phonon scattering was reported for graphene-field-effect devices fabricated on LAO/STO and gating with oxide interface as a back-gate or side-gate, enabling the observation of intrinsic pseudospin quantum transport signature of graphene up to room temperature.[17] This finding has opened up prospects of future works in order to understand unique interaction between these two-dimensionally confined electronic systems. Therefore, understanding the nature of electric-field modulation of carrier transport, ideally by using either side as an electrical gate for the other, is of fundamental importance in developing novel multifunctional devices.

Control of charge carriers via electric-field gating has been widely employed in LAO/STO-based interfacial field-effect transistor (FET) devices through different gating configurations such as a global back gate,[5,7,18,19] global top- and side-gates,[20-23] local top- and side-gates,[14,15,22,24] and electrolytic top-gates.[25] Among all of these gating options, global back gating has some disadvantages due to various properties of STO (e.g., high dielectric permittivity at low temperature), which can complicate the carrier tuning in the nanoscale devices. Other effects are believed to play an important role, including intrinsic structural phase transitions,[26-28] locally ordered structural domains,[29,30] and oxygen vacancies.[31] Use of in-plane side-gates in devices created by c-AFM lithography can introduce other limitations such as current leakage. The top-gating approach with deposited metal electrodes or electrolytes show robust tunability with reduced leakage, since the thin top layer of LAO serves as an excellent dielectric medium.[20,25,32,33] Employing graphene as a top-gate, we can take advantages of its unique properties such as high carrier mobility, high chemical stability, and easily configured nanoscale geometry. Moreover, one could envision exploring novel physics of nanoscale field-effect devices where nanoscale graphene and oxide-based conducting elements are fabricated in a close proximity.

Here we define several reconfigurable nanoscale devices in G/LAO/STO heterostructures and investigate electric-field control of charge transport at the LAO/STO interface using an in-situ patterned graphene structure as a top gate. Unique transport properties of an oxide interface are observed at mesoscopic transport regimes, suggesting highly useful gating options of patterned graphene as a local gate for nanoscale devices. Results are discussed through top-gate dependent $I-V-$curves and magnetoconductance for each device.

## RESULTS AND DISCUSSION

Electric-field control of interfacial conductance is investigated on using several types of field-effect devices fabricated on G/LAO/STO heterostructures. The first two types of devices (Devices A and B) are structured in a typical 3-terminal (3-t) arrangement in which two ends of the interfacial nanowire serves as source (S) and drain (D) and the electrically isolated graphene is used as a top gate (TG) for modulating the electrical transport at the LAO/STO interface at room temperature. The other three types of devices (Devices C, D, and E) are 5-t Hall bars in which 4-t probes of interfacial nanowire device is tuned with a patterned graphene structure as a local top-gate at cryogenic temperatures.

We first investigate the conductance tunability in Device A, shown in Figure **1**(a). A 3-t nanowire is fabricated at the LAO/STO interface, both underneath a graphene-covered region and bare LAO region. The nanowire segment directly underneath the graphene region, as indicated by the red arrow in Figure **1**(a), serves as the active region of the device. A cross-sectional view in the vicinity of the active region is shown schematically in Figure 1(b), where electrons accumulate locally at the interface during the nanowire-writing process. The 2-t conductance between the source electrode $S$ and drain electrode $D$ is monitored while writing the nanowire (Figure **1**(c) and subsequent repeated application of top-gate $V_{tg}$ applied to the graphene (Figure **1**(d). When the graphene top gate is grounded, the device exhibits negligible conductance. When a voltage is applied to the top gate, the nanowire conductance first increases slowly and then rapidly at $V_{tg} \sim 6$ V. Such field-effect gating behavior is repeated several times both in forward and reverse bias conditions, as shown for 12 cycles in Figure **1**(c). Three parametric plots of hysteretic conductance versus top-gate voltage $G(V_{tg})$, for cycles 5, 9, and 12, are shown in Figure **1**(d). A linear fit to the conductance rise indicates a threshold voltage of $V_{th} = 6\ V$ for transition to a conducting state in this particular device (see inset of Figure **1**(d)). It is worth noting that there is negligible conductance hysteresis for any of the gate-sweeping cycles, despite the relatively large top-gate voltage. Hysteresis-free gating could be useful for potential applications in oxide-based electronic devices. Therefore, this observation demonstrates a robust field-effect gating of LAO/STO interfacial devices. The operation of the device is limited only by the gradual decay of the nanowires that lead to the active device, not by the graphene active area itself. The conductance decay can be halted by placing the device in a vacuum environment.[34]

A second class of device, Device B, modifies the geometry so that the nanowire is written only in close proximity to, but not underneath, the graphene. Figure 2(a) shows a plan-view device structure and close-up view in the vicinity of the graphene edge. A side view of the separation region between graphene edge and nanowire, indicated by a red arrow in the plan-view image, is shown schematically in Figure 2(b). In this geometry of the device, writing induces interfacial charges, which are laterally separated from graphene edge. The separation of the nanowire to the graphene is estimated to be $100 \pm 10$ nm. Unlike Device A, discussed earlier, Device B shows a non-zero conductance ($G = 40$ nS) immediately after writing a nanowire between electrodes $S$ and $D$, as indicated by the sharp jump of conductance at 200 s in Figure 2(c). Following the creation of the nanowire, $V_{tg}$ is cycled repeatedly while monitoring the conductance, as shown for 5 cycles in Figure 2(c). Two of the complete cycles, as specified by 2 and 3 are also shown in Figure **2**(d).

In Device B we are able to completely switch -on and -off the nanowire conductance, even though the channel is as far as 100 nm away from the edge of graphene. This proximity-induced field effect resembles the metastable writing that occurs when performing c-AFM lithography on LAO/STO.[34] In a typical c-AFM writing process, a positively-biased AFM tip acts as a source for protonating the surface in a humid environment, resulting in electrons being attracted to the oxide interface. When the tip bias is reversed, the tip removes the $H^+$ and restores the original surface, restoring the initially insulating interface. Here, graphene plays a role that is analogous to the AFM tip, and achieves local hysteretic control of the metal-insulator transition at the LAO/STO interface. The suppressed hysteresis effect in the device made underneath graphene may be caused by graphene gate-induced adsorption and desorption of surface ions. These effects can be completely suppressed when wires are written under vacuum or dry environment.

We next investigate local modulation of the metal-insulator transition in a 4-t LAO/STO nanowire device, created in close proximity to a graphene top gate (Device C). The device architecture and measurement scheme are shown schematically in Figure 3(a). A nano-ribbon-like graphene structure is patterned in situ by anodic etching[17,35,36] of large-area graphene in a predefined path across the G/LAO/STO device. A 4-t nanowire device is created at close proximity to the graphene layer. The active part of the device, shown in Figure 3(c), is separated by $\sim 100$ nm from the edge of the graphene. Four-terminal conductance of the nanowire is monitored during device writing (Figure 3(d)). The conductance slowly decays (typical decay time is $1 - 2$ hours) due to exposure of the surface to ambient atmosphere, but such devices are known to be indefinitely stable under vacuum/cryogenic conditions.[37] This particular device was transferred within 5 minutes to the cryostat and cooled to $T = 2$ K in order to investigate its gate-tunable properties.

$I - V$ curves are measured at different top-gate voltages (Figure 3(e), (f)). Initially, the nanowire is highly conductive. It can be gradually switched to an insulating state by decreasing the top-gate voltage from $V_{tg} = -180$ mV to $V_{tg} = -215$ mV and switched back to the conducting state by increasing the voltage without hysteresis. The onset of the insulating state can be identified from the gradual change of barrier width through the gradient of source-drain current $I_{SD}$ in the $I - V$ map. More insight can be obtained through the gradual change of nonlinear $I - V$ characteristics in Figure 3(f), which confirms the expected field-induced depletion of carriers before finally becoming insulating. Near the metal-insulator transition, we observe oscillations in the conductance (Figure 3(g)) that become more pronounced at the lowest temperatures (2 K). These oscillations are consistent with gate-controlled conductance quantization observed in other sketched LAO/STO devices.[38-40]

To investigate this effect further, top-gated conductance measurements are performed on a similar device (Device D, Figure 3(c)) in an out-of-plane magnetic field $B$ at $T = 50$ mK. At $B = 0$ T, the nanowire conductance is tuned completely from metallic to insulating while sweeping the top-gate from 0.1 V to −0.1 V, as shown in Figure **4**(a). Similar to the previous results observed at 2 K, the onset of a metal-insulator transition shows pronounced oscillatory features. The peak splitting is apparent at different top-gate voltages, which is clearly visible in the magnetoconductance map in Figure **4**(b). The gate range where pronounced oscillations are observed is highlighted in a close-up map and corresponding conductance profiles in Figure **4**(c) and (d). The conductance peak splits at a critical field of $B_C \approx 1.8$ T, above which the energy difference between the split peaks increases linearly with increasing magnetic field. $B_C$ is much higher than the upper critical field for superconductivity ($B_C = 0.2$ T), which is consistent with the phenomenon of electron pairing outside the superconducting state.[41] Unlike the pre-defined single-electron transistor (SET) devices fabricated in the previous work,[41] the graphene gating effect is so strong that barriers in a nominally conductive nanowire can be formed without intentional creation of insulating barriers. In this respect, graphene greatly extends the range for which complex-oxide-based devices can be gated.

    Next, we investigate the conductance modulation by gating nanowire devices with a pre-defined single barrier in Device E. This single-barrier device, as sketched in Figure **5**(a), is fabricated by creating a narrow (~ 5 nm) junction in a device similar to Devices C and D previously discussed. The formation of the barrier is confirmed via monitoring 4-t-conductance in which the conductance drops sharply to zero as soon as negatively biased tip crosses the nanowire path, as indicated by arrow in Figure **5**(b). The electrical transport properties of this devices is then measured at $T = 2$ K. Figure **5**(c) shows a current-voltage map constructed from a series of $I - V$ curves measured at different top-gate voltages. The nanowire is reversibly tuned from an initially insulating state to a conductive state and back by varying the top-gate voltage between $V_{tg} = 0$ V and $V_{tg} = 0.5$ V. The onset of conduction can be identified from the gradual change of the source-drain current $I_{SD}$ and also evolution of $I - V$ characteristics from nonlinear to linear regime as the gate voltage is increased, as shown in Figure

**5**(d). These results demonstrate the narrowing of the barrier via field-induced accumulation of carriers at the interface, which is in contrast to the field-induced depletion of carriers observed earlier.

Finally, we create a double-barrier nanowire device (Device F), schematically shown in Figure **6**(a). Unlike the single-barrier device (Device E), the barriers in Device F are written while writing the main channel by alternating the tip-voltage between $V_{tip} = +15$ V and $V_{tip} = -5$ V. The formation of barriers can be identified by sharp decrease of the conductance of the device, as shown in Figure **6**(b). We first examine the device tunability by measuring the conductance (at $T = 50$ mK) as a function of top-gate voltages $V_{tg}$ in the range $-50$ mV to $100$ mV (Figure **6**(c)). The conductance switches on at $V_{tg} = -35$ mV and increases with increasing $V_{tg}$. At $B = 0$ T, the conductance displays several sharp resonance-like peaks and broader dips over the entire conducting regime. These resonance features weaken considerably at higher magnetic fields. A series of $I - V$ curves are measured over an entire tuning regime. The differential conductance $G = dI/dV_{4t}$ is extracted numerically from each $I - V$-curve at different top-gate voltages $V_{tg}$. The resulting gate-dependent conductance maps at $B = 0$ T and $B = 8$ T are shown in Figure **6**(d). Overall, the behavior is consistent with single-electron transistor devices for which electron pairing outside the superconducting regime was observed.[42] A strongly enhanced conduction is observed in the superconducting regime at higher bias (Figure 6(c)), and characteristic conductance oscillations are observed at all magnetic fields (Figure 6(d)). The resistance in the superconducting regime is non-zero (see Figure S4 in Supporting Information) which has been attributed to gapless quasiparticles.[43] Figure **6**(e) shows the evolution of the zero-bias-peak with increasing field, which is a similar to the measurement shown in Figure **4**(b). Although several fine features are weakly resolved, most of the resonance-like peaks undergo linear Zeeman-like splitting above a critical field $B_c = 3$ T, consistent with the phenomenon of electron-pair breaking.

## CONCLUSION

We have integrated graphene with LAO/STO heterostructures and used the graphene layer as a local top-gate for nanoscale devices fabricated at the LAO/STO interface. A variety of nanoscale devices are created by reconfiguring the LAO/STO conducting region using c-AFM lithography. Under ambient conditions, field-effect tuning of the interface conduction is robust and hysteresis-free for the devices fabricated underneath graphene. LAO/STO devices that have a small lateral separation are found to be strongly hysteretic, suggesting a route for reconfigurable gating without the use of c-AFM lithography. Beyond usual top-gate tuning of devices, local control of confined electron gas is achieved in nanowire devices with and without barriers up to the single-electron tunneling regime at low temperatures. Field-dependent measurements reveal the unique intrinsic phenomena of electron-pairing without superconductivity. These results highlight the excellent gating performance of graphene and pave the way for developing novel multifunctional devices consisting of graphene and complex-oxides.

## METHODS

**LAO/STO sample preparation.** LAO/STO heterostructures are fabricated by growing 3.4 unit cell LAO films on $TiO_2$-terminated STO (001) substrates using pulsed laser deposition (PLD).[44,45] The LAO film thickness is controlled by in-situ monitoring of the reflection high-energy electron diffraction (RHEED) spot intensity during the layer-by-layer growth of LAO. The sample is then patterned with 12-electrodes at each active device region, called a "canvas". The electrodes are made using a two-step deposition process: (1) six electrodes to contact the interface are made by ion (Ar+) milling followed by backfilling with 4 nm titanium (Ti) and 25 nm of gold (Au) via sputter deposition process; (2) six top-electrodes, isolated from the LAO/STO interface, are made by depositing 4 nm/25 nm Ti/Au directly on LAO surface. The interface electrodes make electrical contact with the LAO/STO interface, e.g. to make devices using c-AFM lithography, while the top-electrodes are used to make electrical contact to graphene.

**Graphene synthesis, transfer, and patterning.** Large area single-layer graphene is synthesized by atmospheric pressure chemical vapor deposition (APCVD) on ultra-smooth Cu wafers.[46,47] The graphene layer is transferred onto pre-patterned and mildly oxygen-plasma-cleaned LAO/STO substrates using a standard poly-methyl methacrylate (PMMA) assisted wet-transfer procedure.[48] Circular graphene pieces of ~ 20 μm diameter are patterned at each canvas using standard deep-ultraviolet (DUV) lithography and controlled oxygen plasma etching (see Supporting Information, Figure S1). A mask was employed during DUV exposure in order to precisely align the circular graphene piece with the top-electrodes and electrically isolate it from interface. Additional patterning of graphene to create nanoscale shapes (see Figure S2 in Supporting Information) was performed by in-situ anodic etching using c-AFM lithography.[35,36]

**Sample characterization.** An AFM (Asylum MFP-3D) is used in a non-contact mode to examine the surface morphology of the LAO/STO substrate and transferred graphene, which showed smooth morphology without significant contaminants or trapped water molecules. AFM images are acquired in air using doped silicon cantilevers operated in tapping mode. Raman spectroscopy, using a commercial Raman microscope (Renishaw InVia) with 633 nm laser excitation under ambient conditions, reveals single layer graphene with low doping and defect densities, consistent with AFM results (see Supporting Information). Details of Raman spectroscopy results of similar samples and carrier mobility measurements are described elsewhere.[17] Typical room temperature mobility of the graphene is $> 1\times10^4$ cm$^2$/V·s, indicating high quality graphene samples.

**Nanoscale device fabrication.** Graphene/LAO/STO samples are mounted on a chip carrier, and electrical connections are made by Au wire bonding between the chip carrier and contact pads. All interfacial nanowire devices are fabricated using c-AFM lithography, in which a positively-biased tip ($V_{tip} \sim +10\ V$), scanned in contact mode, locally induces a conducting channel (width ~ 10 nm) across the two LAO/STO interface electrodes. These nanowires can be erased and switched to an insulating state, as required, by scanning a negatively-biased tip ($V_{tip} = -10$ V) across the nanowire. A narrow junction or a nanoscale-insulating barrier is created on the wire by moving a negatively-biased tip across an existing nanowire. The typical width of the barrier, as determined by the resistance profile during the erasing process of conducting path at room temperature, is 5 nm for $V_{tip} = -1$ V. The width of the nanowire or a junction is limited by the AFM-tip area of contact, tip voltage $V_{tip}$, writing speed, and other environmental conditions.[49]

**Transport characterization.** Electric-field control of interfacial conductance is investigated using several types of field-effect devices fabricated on graphene/LAO/STO heterostructures. Room temperature conductance measurements are performed using standard lock-in techniques at a reference frequency of 33 Hz. The sample is grounded for each measurement. For recording gate-dependent $I-V$-curves at low-temperatures ($T \geq 2$ K), standard lock-in technique was used at a reference frequency of $f = 3.46$ Hz and ac amplitude of 20 µV. Low-temperature measurements were performed in a helium cryostat (Quantum Design PPMS).

## ACKNOWLEDGMENT

The authors would like to thank Shicheng Lu and Anil Annadi for their help in low-temperature transport measurements. We gratefully acknowledge support from the Office of Naval Research (N00014-13-1-0806, N00014-16-1-3152) (C.B.E. and J.L.) and the AFOSR FA9550-15-1-0334 (C.B.E.), the National Science Foundation Nos. DMR-1234096 (C.B.E.) and DMR-1104191 (J.L.).

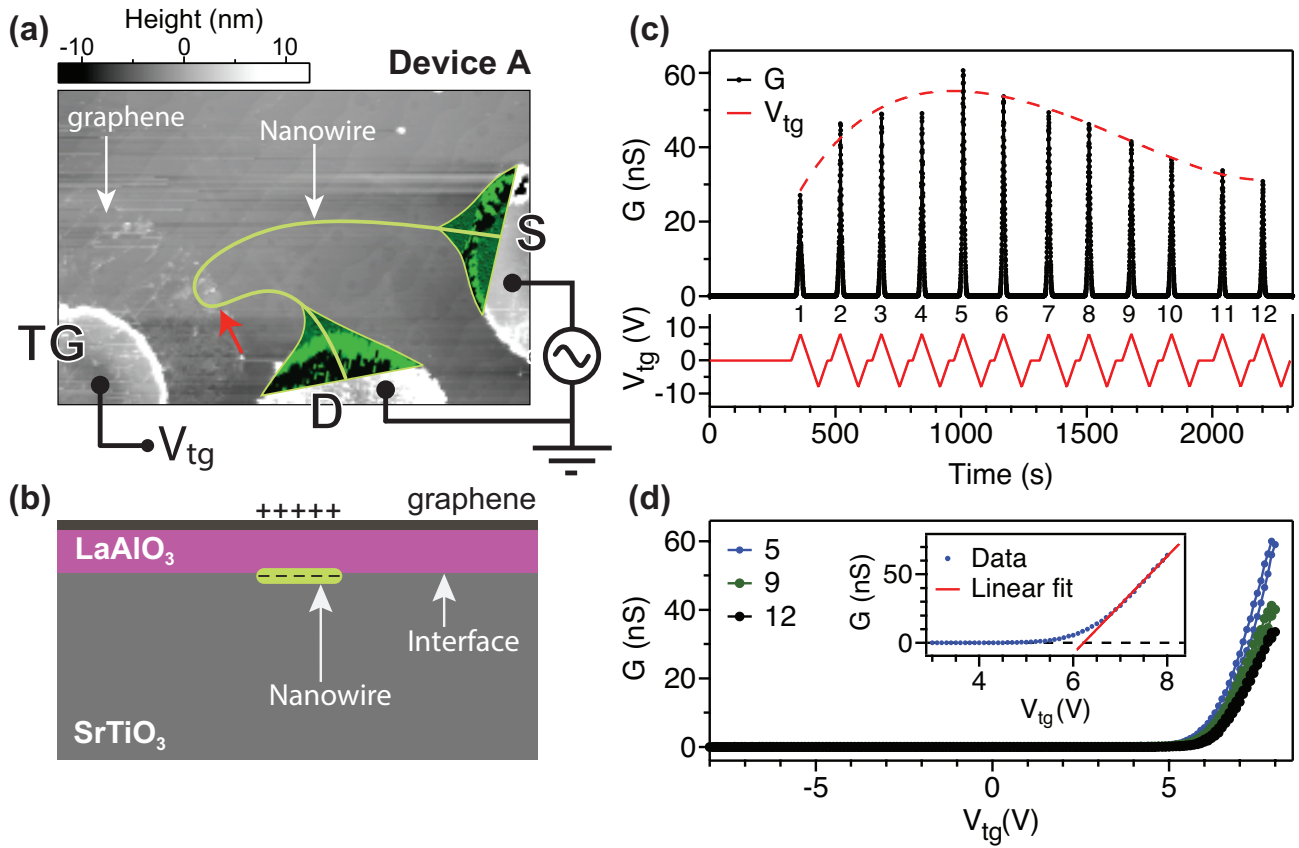

Figure 1. Device A (a) Tapping mode AFM image (with enhanced contrast) of a 3-t nanowire device fabricated on a G/LAO/STO heterostructure. Two interface-connected electrodes are used as source (S) and drain (D). A top-electrode, isolated from the LAO/STO interface, contacts the graphene and is used as a top-gate (TG). An interfacial nanowire, fabricated using c-AFM lithography, is indicated by a light green line; funnel-shaped dark-green regions represent virtual electrodes. (b) Schematic cross section of the device, showing gate-induced electrons that have accumulated locally at the interface. (c) Conductance ($G$) monitoring of the nanowire (top panel) during application a sequence of 12 top-gate pulses (bottom panel) after writing the device at room temperature. (d) Simultaneously measured nanowire conductance as a function of top-gate voltages ($V_{tg}$) during each cycle of top-gate tuning (only three gating cycles 5, 9 and 12 are shown). Inset shows a close-up at the onset of conduction (cycle 5). A linear fit to the onset of conductance rise yields an activation barrier of 6 V for this device.

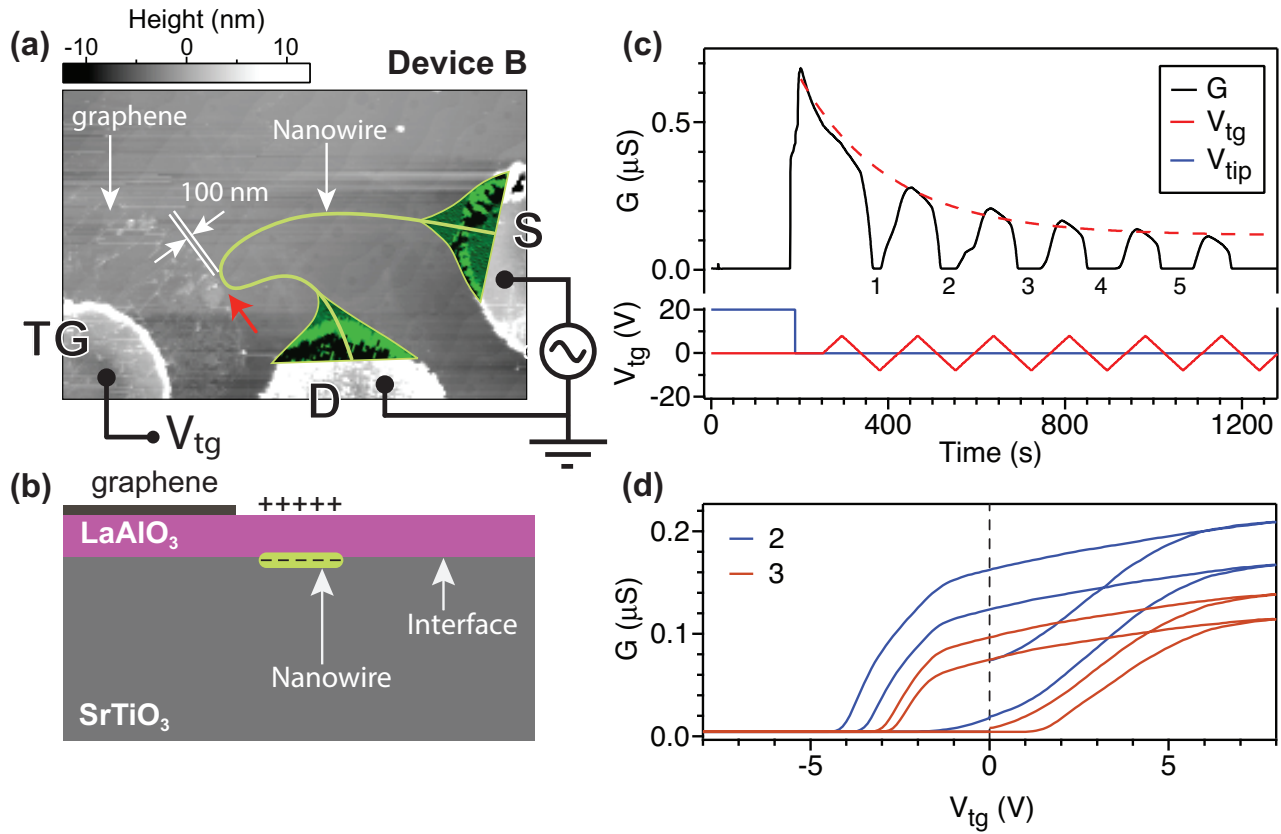

Figure 2. Device B (a) Tapping-mode AFM image of a 3-t nanowire device, with nanowire path overlaid. Interfacial nanowire, fabricated in close proximity of graphene (100 nm away from the graphene edge) by using c-AFM lithography, is indicated by light green lines. Green funnel-shaped regions are virtual electrodes to connect nanowire with electrodes. (b) Cross section view across the written nanowire at the position shown by red arrow in (a). (c) 2-t conductance of nanowire vs. time following writing and during a sequence of alternating voltages applied to the top-gate. (d) Nanowire conductance hysteresis as a function of top-gate voltages $V_{tg}$ during two of the cycles of top gating. A hysteresis-like shift in conductance at each cycle of forward and reverse gate tuning is caused by metastable charge state at the surface, described in the text. The conductance gradually decreases as the carriers decay at the interface after longer exposure to ambient conditions after writing. All measurements including c-AFM lithography and gate tuning of the devices are performed at room temperature under ambient conditions.

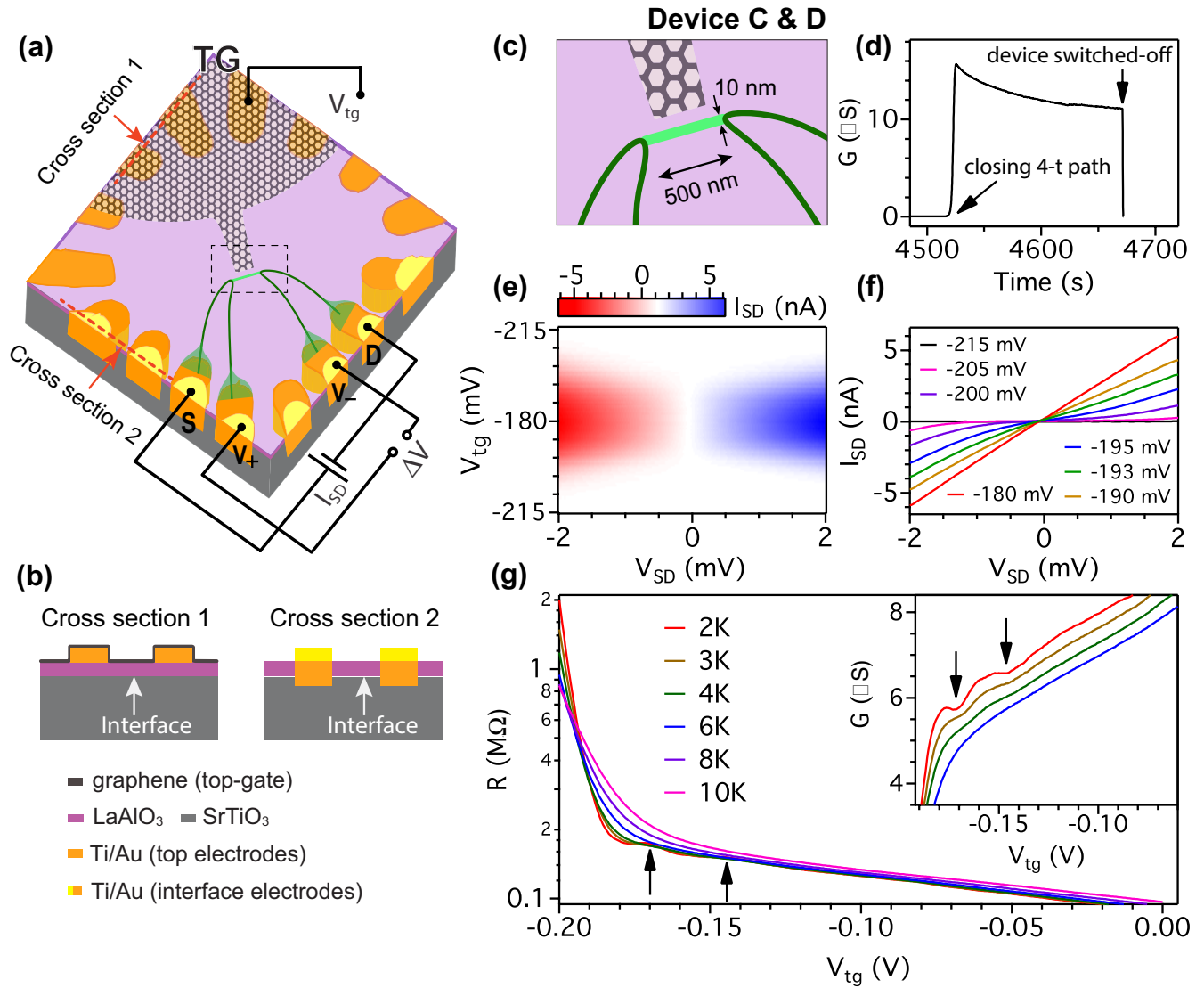

Figure 3. Device C. (a) Schematic view and electrical characterization scheme of the interfacial nanowire device on G/LAO/STO. A ribbon-like graphene structure with nominal width of 500 nm is patterned and a nanowire at the LAO/STO interface is fabricated near (∼ **100** nm) the graphene edge using c-AFM lithography. The active part of the device is shown with light green line and the 4-terminal interconnections are shown with dark green lines. (b) Cross-sectional views of the device structure from the two sides of the device. Cross section 1 shows top-gating geometry with graphene. Cross section 2 shows the electrodes connected to the interfaces, which are used for fabricating nanowires or nano-junctions. (c) Schematic view of the active part of the nanowire device, as indicated by the dashed rectangle in (a), denoted by Device C & D. (d) Conductance monitoring during nanowire writing by c-AFM lithography in which the conductance jump indicates the formation of the active part of the device. (e) $I − V$-mapping as a function of the top-gate voltage ($V_{tg}$), showing a reversible transition

from metallic to insulating phases. (f) Selected $IV$-curves as a function of $V_{tg}$, showing the onset of conducting phase. (g) Gate dependent 4-t resistance ($R$) of the device measured at different temperatures. The resistance shows an oscillatory feature before switching-off to the insulating phase and the feature weakens significantly as the sample is warmed up beyond 6 K. The weakening of the features is clearly visible in the vertically-shifted conductance ($G$) plots shown in the inset. Arrows mark the oscillation minima (maxima) of the conductance (resistance).

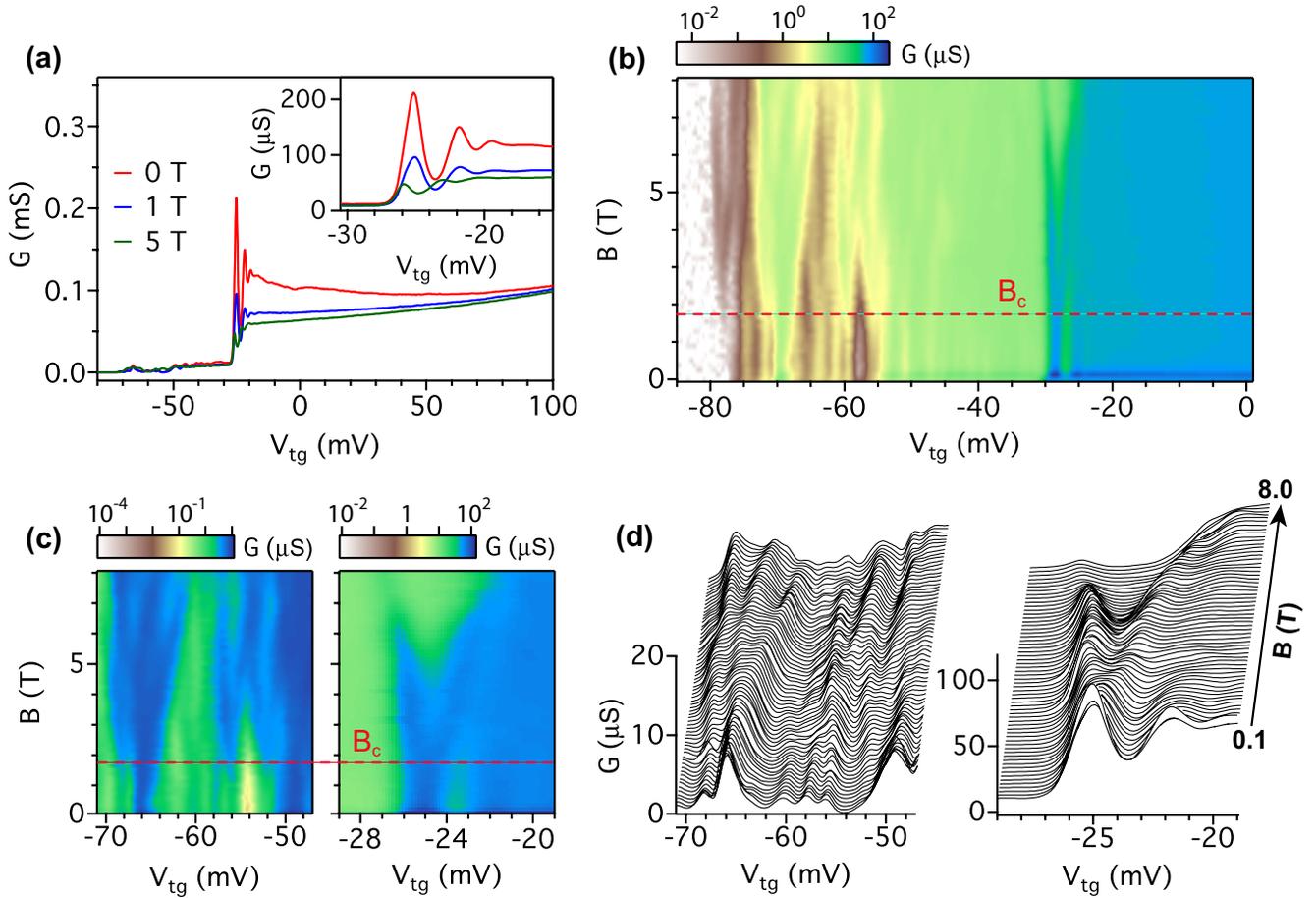

Figure 4. Device D. Schematic of device shown in Figure 3. (a) Gate dependent 4-terminal conductance ($G$) with and without applied out-of-plane magnetic fields. Inset shows a close up view of the conductance oscillations at $V_{tg} \sim 20$ mV in which the peak splits at higher fields. (b) Gate dependent conductance mapping during continuous sweeping of the magnetic field from zero to 8 T. The oscillatory features evolve into more peaks at different energies at higher fields, indicating a clear signature of peak splitting. (c) & (d) Close-up views of the mapping and corresponding line plots, showing the onset of peak splitting at a critical field of $B_c \approx 1.8$ T.

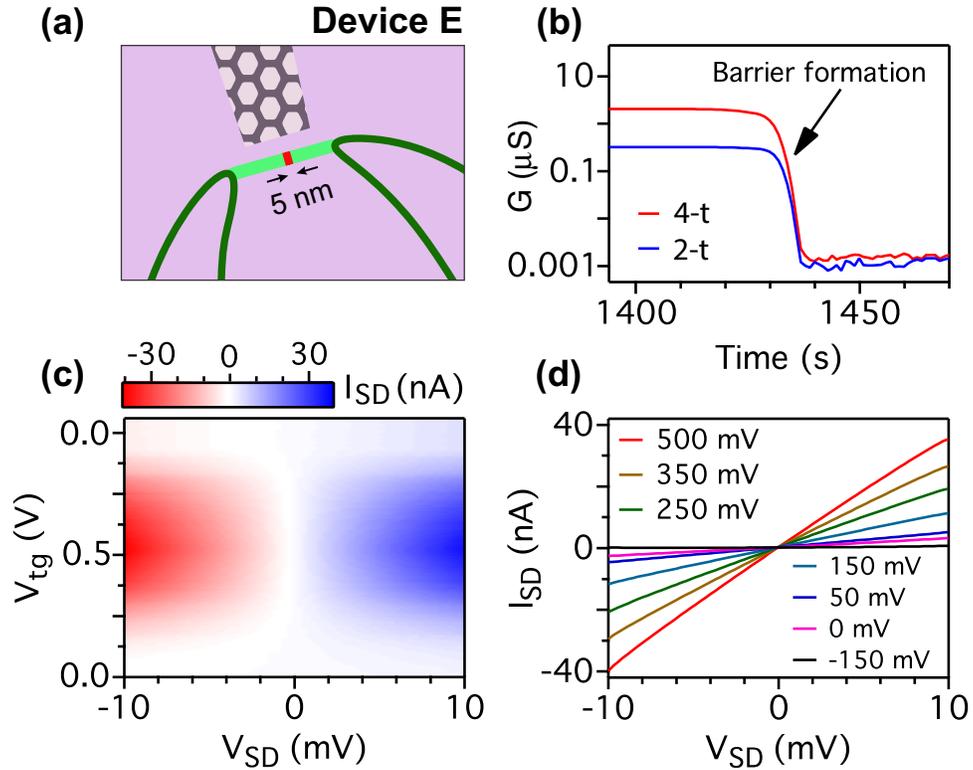

Figure 5. Device E. (a) Schematic view of an active, a nanowire with a single-barrier. (b) Conductance monitoring during a barrier formation, which is identified by the sharp decrease of conductance as soon as the 4-t conducting path is cut. After the device is fabricated at room temperature, its electrical properties are measured at 2 K. (c) Source-drain $I-V$-mapping of the device over a wide range of top-gate voltages ($V_{tg}$), showing a reversible transition from insulating to metallic phases. (d) Selected $I-V$-curves as a function of $V_{tg}$, showing an onset of transition to conducting regime.

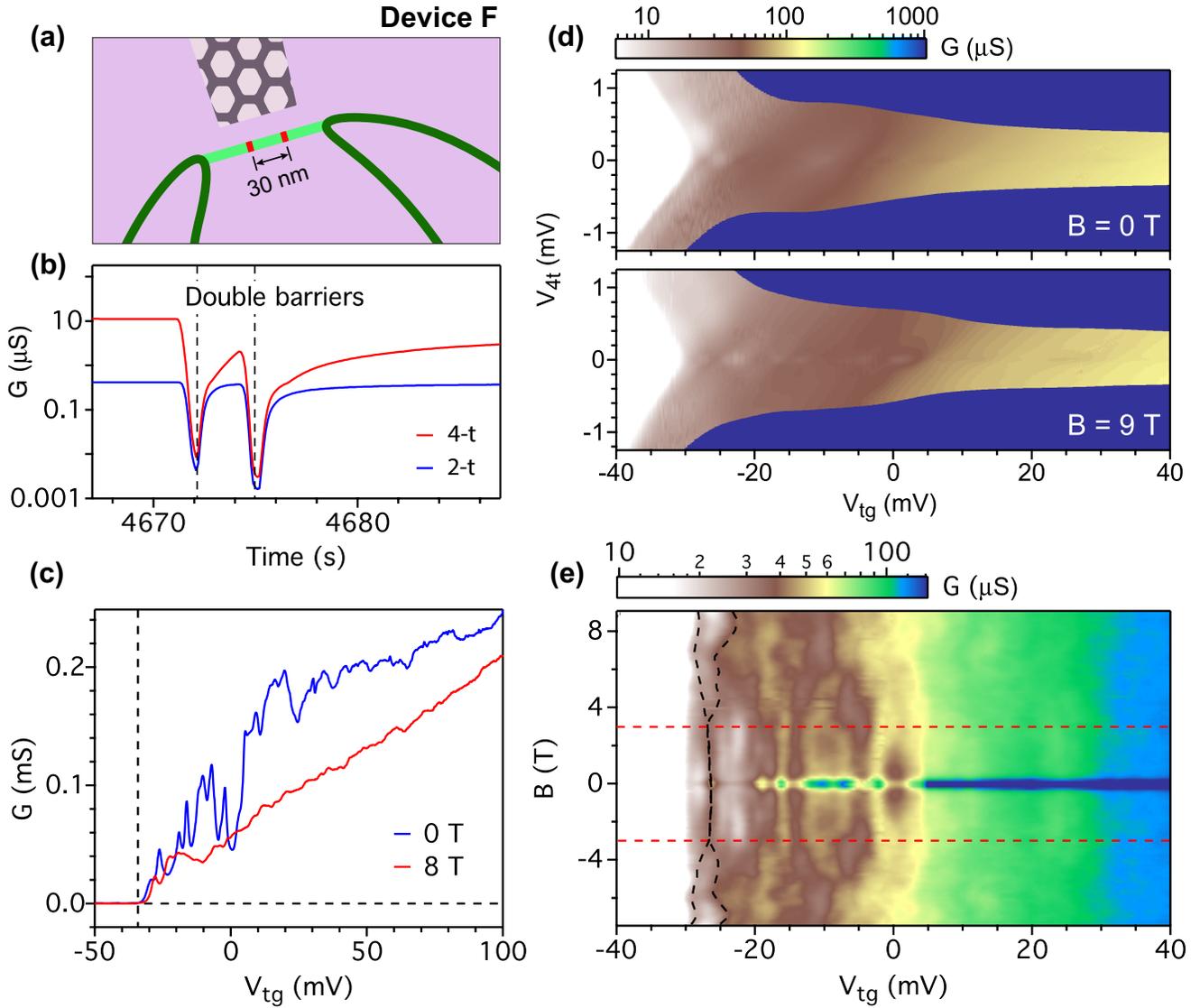

Figure 6. Device F. (a) Schematic view of an active region of the double barrier nanowire device. (b) Four-terminal (4-t) and 2-t Conductance monitoring of the device during fabrication of double barriers (double junctions) followed by nanowire device. The conductance drops to nearly 0 as the barriers are formed. The subsequent gradual increase in the conductance is caused by thermally excited carriers crossing the narrow barriers at room temperature. (c) Gate dependent conductance at zero-field and 8 T during initial test of device tunability at 50 mK. (d) Gate dependent conductance mapping of the device at 0 and 8 T. Sharp conductance peaks and weakly formed Coulomb diamonds are visible near zero bias, indicating single-electron tunneling across the quantum-dot-like structure. (e) Continuous sweeping of the magnetic field during conductance tuning of the device. Zero bias peaks split into multiple peaks and smooth out at the fields higher than a critical field of $\boldsymbol{B_c} \approx \boldsymbol{3}$ T, indicated by horizontal red dashed lines. Thin black dotted lines follow the splitting as a guide to the eye.

*Supporting Information:*

**Graphene-Complex-Oxide Nanoscale Device Concepts**

Giriraj Jnawali,[1,2] Hyungwoo Lee,[3] Jung-Woo Lee[3], Mengchen Huang,[1,2] Jen-Feng Hsu,[1,2] Bi Feng,[1,2] Rongpu Zhou,[1,2] Guanglei Cheng,[1,2] Brian D'Urso,[1,2] Patrick Irvin,[1,2] Chang-Beom Eom,[3] and Jeremy Levy[1,2†]

[1]*Department of Physics and Astronomy, University of Pittsburgh, Pittsburgh 15260, USA*
[2]*Pittsburgh Quantum Institute, Pittsburgh, PA 15260, USA*
[3]*Department of Materials Science and Engineering, University of Wisconsin-Madison, Madison, WI 53706, USA*

† Correspondence and requests for materials should be addressed to J.L. (jlevy@pitt.edu).

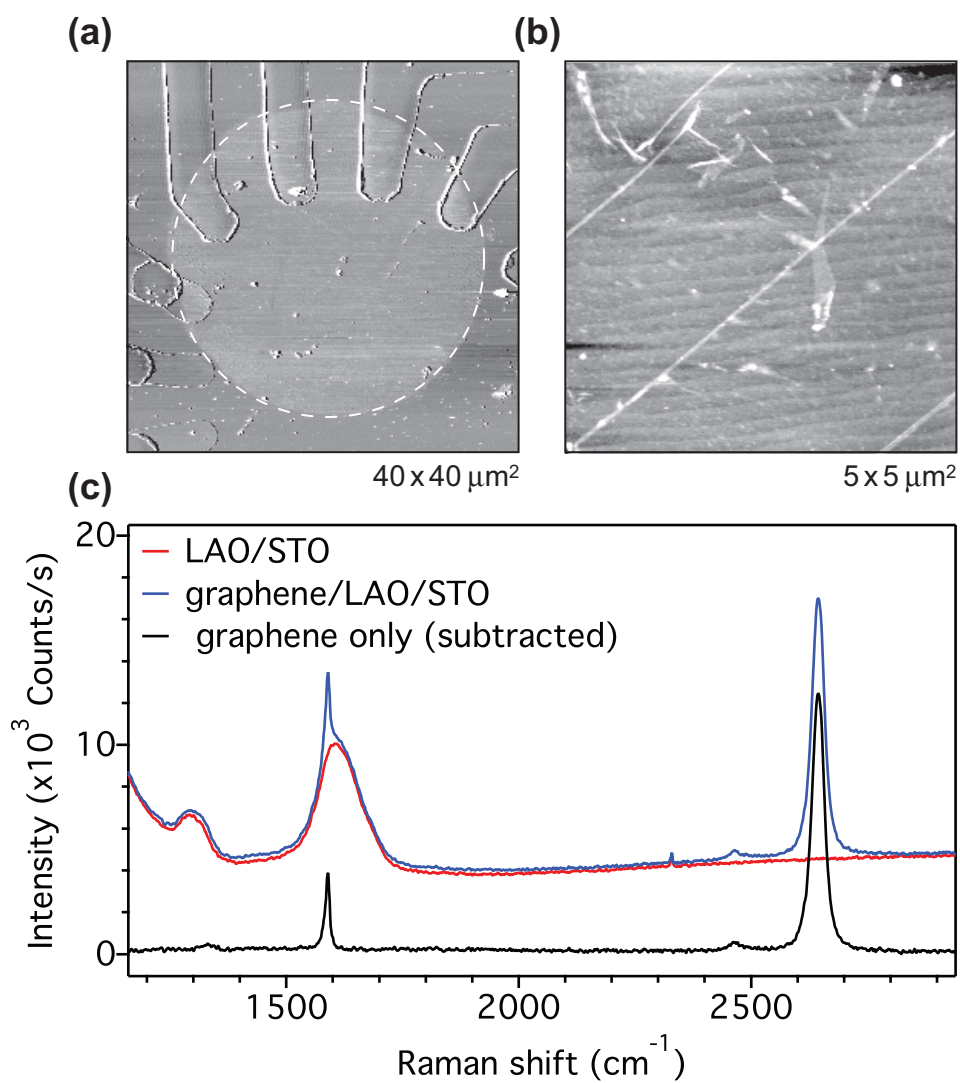

Figure S1. Surface morphology of graphene/LAO/STO (G/LAO/STO) by AFM and Raman spectroscopy. (a) Tapping mode AFM height images of G/LAO/STO samples after transferring graphene onto the LAO/STO substrate. The graphene is patterned is patterned into a circular shape (indicate by dashed white lines) via oxygen plasma etching process. (b) Close-up AFM image acquired on the G/LAO/STO region, showing vicinal atomic steps of LAO/STO. **(c)** Raman spectra measured on bare LAO/STO and on regions covered with graphene. The subtracted spectrum is also shown (black solid line) to distinguish Raman features only from graphene. Subtracted spectrum shows G-peak at 1585 cm$^{-1}$ and 2D-peak with a single Lorentzian profile at 2660 cm$^{-1}$. The intensity ratio of 2D-peak to G-peak is almost 3. All of these features confirm a single layer graphene with low-defect density.

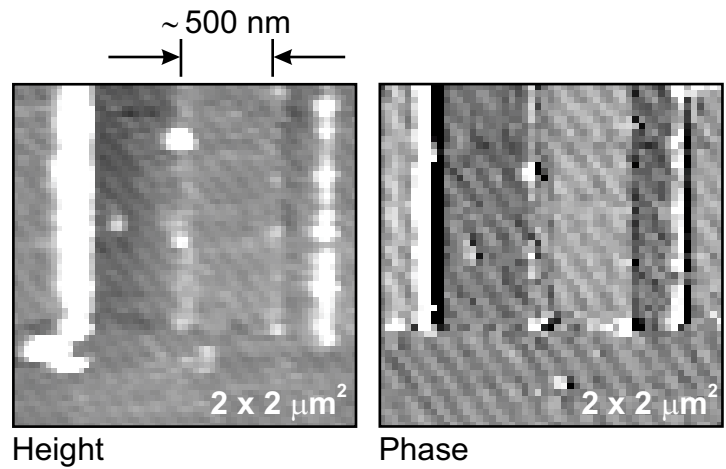

Figure S2. Patterned graphene structure for top-gate operation. Tapping mode AFM height (Left) and phase (Right) images of a ribbon-like graphene structure used for top-gate operation. Slanted diffused white lines are background noise appeared during low-resolution AFM imaging. Patterning is performed by anodic etching with a biased AFM tip ($V_{tip} \sim 30$ V). Some PMMA residue accumulated after cutting the graphene is carefully removed by contact mode scanning outside the graphene pattern. In particular, the region in front of the graphene edge is cleaned by repeating the scanning process. Typical width of the graphene structure, used for our gating operation and closest to the device edge, is approximately $\sim 500 \pm 50$ nm. The error may be due to the presence of PMMA residue (white particles in height image). The width can be further scaled-down by using a sharper tip (tip diameter $\leq 5$ nm) than the one (10 nm diameter) we have used.

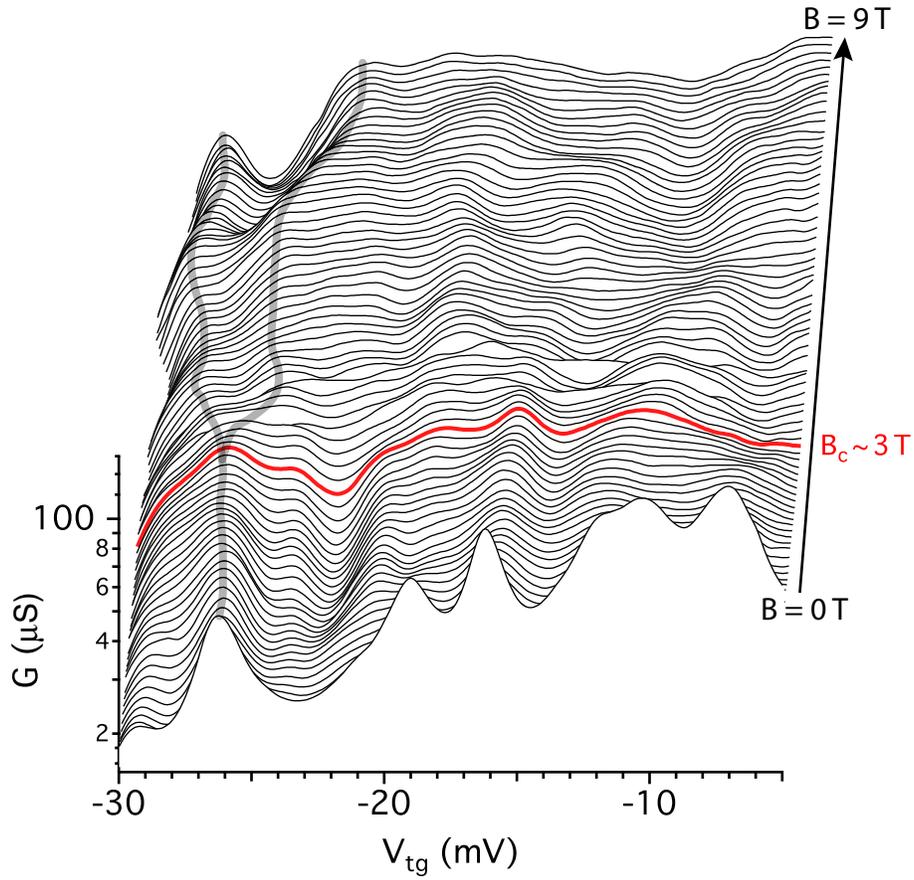

Figure S3. Gate-dependent continuous field-sweep. Continuous sweeping of the field from 0 T to 9 T during conductance tuning of the double barrier nanowire device, Device F. Zero bias peaks split into multiple peaks and smooth out at higher fields ($B > B_c \sim 3$ T). The red curve highlights the onset of splitting. The grey lines follow the peak as a guide to eyes.

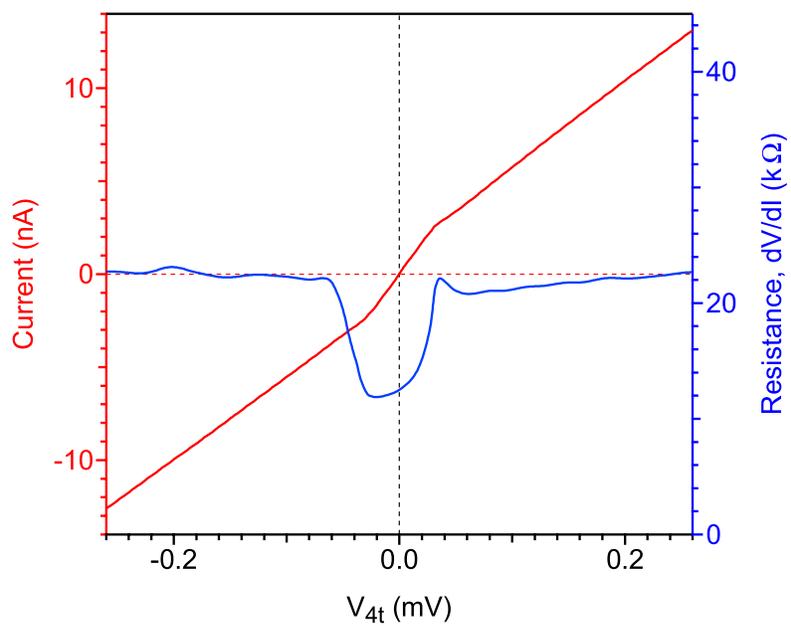

Figure S4. *I-V*-curve and differential resistance of a single interfacial nanowire device, Device F (see Figure 6), fabricated by c-AFM lithography on LAO/STO heterostructure.